\documentclass[
    superscriptaddress,
    reprint, 
    amsmath,amssymb,
    aps,
    prl,
    floatfix,
]{revtex4-2}

\usepackage{graphicx}
\usepackage{xcolor}
\usepackage{siunitx}
\usepackage{bm}
\usepackage[mathlines]{lineno}
\usepackage{mathtools}
\usepackage[colorlinks=true,linkcolor=blue,citecolor=blue]{hyperref}
\usepackage[capitalise]{cleveref}
\usepackage{csquotes}

\AddToHook{cmd/appendix/before}{\crefalias{section}{appendix}}
\AddToHook{cmd/appendix/before}{\crefalias{subsection}{appendix}}

\DeclarePairedDelimiter\ket{\lvert}{\rangle}
\def\({\left(}
\def\){\right)}

\begin{document}
\title{Parametrically induced strong coupling between a superconducting quantum circuit and a solid-state spin ensemble}

\author{Alejandro E.~Baptista}
\email{aeb7@illinois.edu}
\author{Jinwoong Kim}
\author{Sonia Rani}
\author{Xi Cao}
\affiliation{Department of Physics, The Grainger College of Engineering, University of Illinois at Urbana-Champaign, Urbana, IL 61801, USA}
\author{Wolfgang Pfaff}
\email{wpfaff@illinois.edu}
\affiliation{Department of Physics, The Grainger College of Engineering, University of Illinois at Urbana-Champaign, Urbana, IL 61801, USA}
\affiliation{Materials Research Laboratory, The Grainger College of Engineering, University of Illinois at Urbana-Champaign, Urbana, IL 61801, USA}
\affiliation{Holonyak Micro and Nanotechnology Lab, The Grainger College of Engineering, University of Illinois at Urbana-Champaign, Urbana, IL 61801, USA}
\affiliation{National Center for Supercomputing Applications, University of Illinois at Urbana-Champaign, Urbana, IL 61801, USA}

\begin{abstract}
Efficient quantum state transfer between superconducting circuits and solid-state spins would unlock high-coherence quantum memories for superconducting quantum processors.
We demonstrate dynamically controlled strong coupling between a Josephson circuit and a rare-earth spin ensemble.
Using a parametric pump, we realize on-demand coupling of several MHz, which will enable faithful state transfer between quantum circuits and spins.
Our architecture enables quantum control of spin ensembles, and paves the way for hybrid memories with coherence far beyond those of superconducting circuits alone.
\end{abstract}
\maketitle

Superconducting circuits are a leading platform for quantum information processing. 
They have recently met critical milestones such as below-threshold quantum error correction \cite{acharya_quantum_2025a, he_experimental_2025} and promising signatures of quantum advantage \cite{kim_evidence_2023, googlequantumaiandcollaborators_observation_2025}.
Their performance, however, remains limited by available coherence: dielectric losses, coupling to two-level systems, and other intrinsic loss mechanisms restrict storage times to the millisecond range \cite{somoroff_millisecond_2023, ganjam_surpassing_2024a, bland_millisecond_2025}.
Augmenting superconducting processors with passive quantum memories
would substantially reduce the overhead for fault-tolerance~\cite{gouzien_factoring_2021,mundada_heterogeneous_2026} and enable applications with inherent latency, such as probabilistic quantum communication \cite{briegel_quantum_1998, sangouard_quantum_2011, azuma_quantum_2023, gu_hybrid_2025}.

A particularly attractive memory candidate is provided by solid-state spin ensembles: 
They offer some of the longest known solid-state coherence times, up to several hours~\cite{wang_nuclear_2025a}; this is far beyond the  tens of milliseconds of the best available memory, the 3D cavity~\cite{reagor_quantum_2016a, milul_superconducting_2023a}. 
They further have atomic-like and well-characterized energy-level structures, and can have ground-state transitions in the GHz range that are intrinsically compatible with superconducting circuits~\cite{ledantec_twentythreemillisecond_2021, alexander_coherent_2022, wang_nuclear_2025a, chen_coupling_2016}.
Additionally, spin ensembles are promising for dense multi-mode storage~\cite{julsgaard_quantum_2013}.
Already, storage and retrieval of classical states on the 100-ms scale as well as random-access have been demonstrated in hybrid circuits coupled to spin ensembles~\cite{grezes_multimode_2014a, ranjan_multimode_2020, osullivan_randomaccess_2022}.
An open question remains, however, how quantum states can be faithfully encoded and retrieved.

\begin{figure}
    \centering
    \includegraphics[]{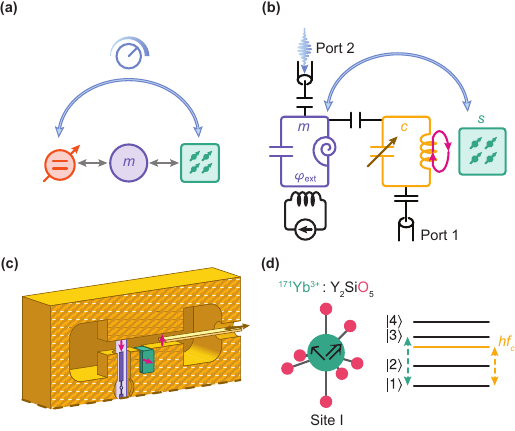}
    \caption{
    \textbf{Experiment design.}
    (a) Our goal is on-demand, tunable strong coupling via a mixer (purple), $\hat{m}$.
    (b) Lumped element circuit to illustrate our tunable coupling experiment. 
    (c) 3D cross-section sketch of the cavity used to implement the architecture.
    The cavity (yellow), $\hat{c}$, hosts a chip containing a SNAIL coupler (purple), $\hat{m}$, spin ensemble crystal (green), $\hat{s}$, and dielectric slab (light yellow) for cavity frequency tuning.
    Red arrows show the oscillating fields for the differential cavity mode that couple capacitively to the SNAIL and inductively to the spins. 
    A flux coil is inserted near the SNAIL loop for magnetic flux tuning. 
    (d) Spin system: $^{171}\mathrm{Yb}^{3+}$:YSO; cavity frequency is close to the site I clock transition between $\ket{1}$ and $\ket{3}$ at 2.87~GHz \cite{tiranov_spectroscopic_2018}.
    }
    \label{fig:concept}
\end{figure}

Efficient state transfer between circuits and spins requires a tunable coupling between them.
Most recent proposals address this need with absorption and re-emission of wavepackets through resonators that are inductively coupled to the spins~\cite{julsgaard_quantum_2013, wen_addressing_2025,bernad_analytical_2025}.
Reported conversion efficiencies between propagating microwaves and spins, however, have been too low for faithful quantum state encoding and retrieval~\cite{osullivan_randomaccess_2022,ranjan_multimode_2020}.
To create a deterministic quantum memory, a more direct approach would be to realize a quantum gate in the form of an on-demand excitation swap between superconducting circuit and spins [\cref{fig:concept}(a)].
This approach requires dynamically generated strong coupling between a quantum circuit and a highly coherent collective spin excitation.
Static strong coupling has been observed in multiple experiments~\cite{kubo_strong_2010,schuster_highcooperativity_2010,amsüss_cavity_2011,probst_anisotropic_2013};
a critical question remains, however, how to construct a hybrid quantum device that realizes strong coupling dynamically and enables quantum state transfer in analogy to memories with superconducting cavities or mechanical oscillators~\cite{hann_hardwareefficient_2019, li_cascaded_2025}.

Here, we propose and realize a suitable hybrid device architecture and demonstrate dynamically controlled strong coupling between a superconducting quantum circuit and a rare-earth spin ensemble.
We couple a nonlinear Josephson element to a microwave bus cavity that is, in turn, dispersively coupled to an ensemble of $^{171}\mathrm{Yb}^{3+}$ spins doped into a $\mathrm{Y}_2\mathrm{SiO}_5$ (YSO) host crystal.
Through a parametric pump, we hybridize the circuit on demand with a collective spin excitation, and observe a normal-mode splitting that is the hallmark of the strong-coupling regime.
Combined with a qubit, the experimentally demonstrated couplings will enable sub-microsecond quantum state transfer.

Our experimental approach is centered around established, superconducting parametric couplers~\cite{niskanen_quantum_2007, allman_rfsquidmediated_2010, zakka-bajjani_quantum_2011} that 
have enabled fast beamsplitter operations between bosonic qubits stored in cavities~\cite{gao_programmable_2018, chapman_highonoffratio_2023}.
Spin ensembles, however, differ from cavities in a few key ways:
First, spins couple inductively rather than capacitively to a circuit.
Second, ensembles are inherently multi-mode; coupling does not occur to an isolated mode, but to a bright mode of $N$ spins, described by a bosonic operator $\hat s = (\sum_j g_j \sigma_-^{(j)})/\bar g \sqrt{N}$, where $g_j$ are the single-spin couplings, and $\bar g$ is the root-mean-square coupling~\cite{wesenberg_quantum_2009}.
Finally, the attainable collective coupling, $ \bar g\sqrt N$, of circuits to high-coherence ensembles is smaller than what can be achieved with cavities, typically on the order of a few MHz~\cite{ranjan_multimode_2020,alexander_coherent_2022}.

These considerations lead to the design shown in \cref{fig:concept}(b).
The ensemble mode $\hat s$ is coupled inductively to a microwave cavity, $\hat c$, that acts as a bus; the cavity is in turn coupled capacitively to a nonlinear coupler, $\hat m$.
All static couplings are dispersive, $g_{ij}/\Delta_{ij} \lesssim 0.1$, so that without pumping the modes interact only through small dressings; $g_{ij}$ and $\Delta_{ij}$ are the static couplings and detunings between modes $i,j$, respectively.
Our central objective is dynamic strong coupling between the coupler and the spin ensemble---a capability that combines naturally with qubits to enable quantum state encoding and decoding.

We implemented this concept in a 3D architecture [\cref{fig:concept}(c)].
The bus mode is formed by a 3-loop, 2-gap aluminum cavity, whose central loop produces a confined and homogeneous magnetic field over the volume of the spin crystal \cite{angerer_collective_2016, chen_coupling_2016}.
We tune the cavity frequency \emph{ex situ} into the dispersive regime relative to the spin transition using a dielectric slab [\cref{fig:concept}(c)].
The nonlinear coupler is an on-chip LC oscillator containing a Superconducting Nonlinear Asymmetric Inductive eLement (SNAIL) that capacitively couples to a cavity gap and serves as a three-wave mixer \cite{frattini_3wave_2017}; for conciseness, we simply refer to this mixer mode $\hat m$ as `SNAIL' in what follows.

The spin ensemble is hosted in a crystal of $^{171}\mathrm{Yb}^{3+}$:YSO  doped at 10~ppm, which is an effective testbed for our purposes:
This spin system is well-characterized, has relatively high coherence, and is compatible with superconducting circuits \cite{lim_coherent_2018, alexander_coherent_2022}.
The host-crystal anisotropy lifts the hyperfine degeneracy to produce a set of clock transitions spanning 0.3 to 3.2~GHz at zero magnetic field; these transitions are first-order insensitive to magnetic-field fluctuations \cite{tiranov_spectroscopic_2018, ortu_simultaneous_2018, welinski_highresolution_2016}.
They are well matched to superconducting circuits in three respects: their GHz frequencies require no external magnetic field tuning; their magnetic dipole moments are sufficient to expect efficient coupling \cite{tiranov_spectroscopic_2018}; and their reported coherence times of $\sim 10$~ms~\cite{alexander_coherent_2022}, while not the longest available in spins, exceed those of state-of-the-art superconducting qubits.
We target the site I clock transition between ground-state hyperfine levels $\ket{1}$ and $\ket{3}$ at 2.87~GHz.
Finite-element electromagnetic simulations predicted a collective spin-cavity coupling of 1.4~MHz.
We thus placed the bus cavity $\sim 10$~MHz below this transition to place the cavity-spin coupling in the dispersive regime.
Details on the hybrid circuit are presented in \cref{section:exp_set}.

To verify system parameters and investigate the system under the influence of parametric pumps, we used microwave spectroscopy at a temperature of $T = 10\,\text{mK}$.
We performed transmission and reflection measurement on microwave ports attached to the cavity that also allowed for parametric pumping.
External magnetic flux was supplied to the SNAIL through a superconducting coil in the cavity. 
The flux from the coil simultaneously sets the SNAIL frequency and modifies its nonlinear properties.
Full details of the setup and measurement techniques are provided in \cref{section:exp_set} and \cref{section:Meas}.

\begin{figure}
    \centering
    \includegraphics[]{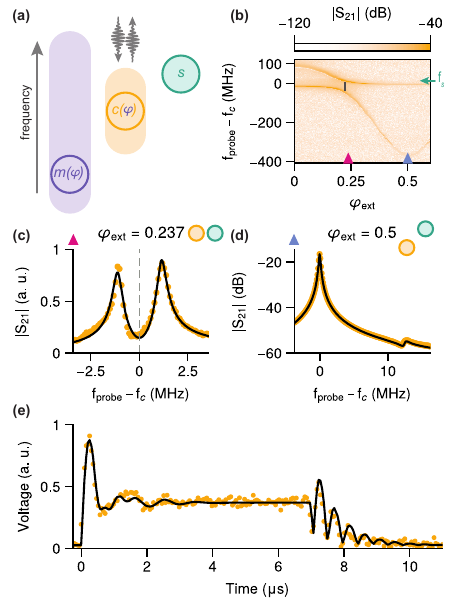}
    \caption{
    \textbf{Mode characterization.}
    (a) Measurement scheme: The cavity was probed as a function of the reduced external flux ($\varphi_\mathrm{ext} = \Phi_\mathrm{ext}/\Phi_0)$ applied to the SNAIL.
    Mode hybridization leads to a modest flux tuning of the cavity mode.
    System is probed by measuring transmission and reflection of the cavity.
    (b) Cavity spectroscopy vs.\ $\varphi_\mathrm{ext}$.
    Gray bar indicates cavity-SNAIL coupling.
    The arrow labeled $f_s$ is at the spin transition frequency.
	(c) Line cut at $\varphi_\mathrm{ext} = 0.237$ (pink marker in (b)).
	(d) Line cut at $\varphi_\mathrm{ext} = 0.5$ (blue marker in (b)).
	(e) Polariton ring-up and ring-down at cavity-spin resonance (gray dashed line in (c)).
	The cavity was driven at $f_c$ with a 7-\textmu s pulse, and response was recorded as function of time.
    The initial overshoot of the ring-down signal is indicative of the non-Markovian spin bath~\cite{putz_protecting_2014, diniz_strongly_2011}.
    }
    \label{fig:flux_spec}
\end{figure}

We first determined the operating point for dispersive coupling through spectroscopy as a function of external flux.
By monitoring the cavity response, we observed avoided crossings between the cavity and other modes in the system, from which we extracted their static couplings and linewidths [\cref{fig:flux_spec}(a,b)].
We measured a static SNAIL-cavity coupling of 27~MHz [\cref{fig:flux_spec}(b)] and a static cavity-spin coupling of 1.1~MHz [\cref{fig:flux_spec}(c)]; 
the spin excitation mode at 2.87~GHz is confirmed explicitly by the ring-down behavior of cavity-spin polaritons~\cite{putz_protecting_2014} [\cref{fig:flux_spec}(e)] and repeated measurements in multiple cavity designs.
The linewidths for the SNAIL, cavity, and spin modes are $\kappa_m = 2\pi\times0.8-1.4$~MHz (varying with flux), $\kappa = 2\pi \times 200$~kHz, and $\gamma = 2\pi \times 460$~kHz respectively, resulting in a cavity-spin cooperativity of $C = 4g_{cs}^2/\kappa\gamma = 48$.
Knowledge of the coupling strengths allowed us then to tune the SNAIL to a strong-dispersive operating point with $g_{cs}/\Delta_{cs} = 0.089$ between the cavity and the spins and $g_{mc}/\Delta_{mc} = 0.098$ between the SNAIL and the cavity.
In this configuration, the spins remained visible in transmission as a small dispersive feature in the cavity response [\cref{fig:flux_spec}(d)].
See \cref{section:Meas} for details.

\begin{figure}[t]
    \includegraphics[]{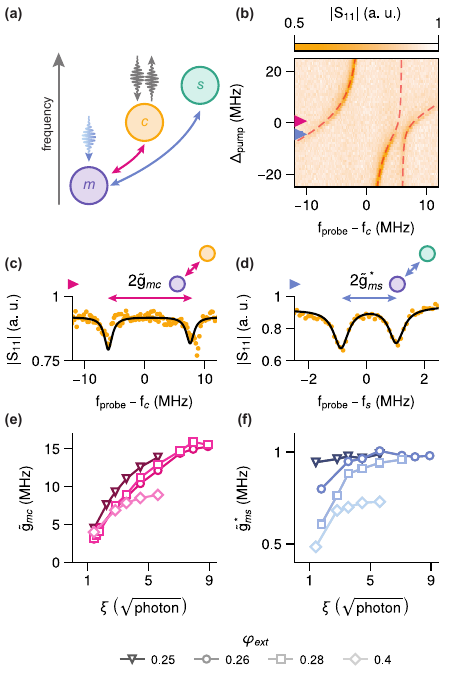}
    \caption{
    \textbf{Parametrically induced strong coupling.}
    (a) At fixed $\varphi_{\mathrm{ext}}$ the SNAIL is driven near SNAIL-cavity (magenta) or SNAIL-spin (blue) difference frequency.
	(b) Cavity spectroscopy in the presence of the pump.
	Dashed lines are simulated frequencies of the 3 coupled modes in the system when the drive is on.
	Here $\Delta_\mathrm{pump}/2\pi = f_{\mathrm{pump}} - \Delta_{mc}/2\pi$ is the difference between the pump frequency, $f_\mathrm{pump}$, and the cavity-SNAIL detuning, $\Delta_{mc}/2\pi$.
	(c) Line cut of the SNAIL-cavity avoided crossing (pink marker in (b)).
	(d) Line cut of the SNAIL-spin avoided crossing (blue marker in (b)).
	(e) SNAIL-cavity coupling, $\tilde{g}_{mc}$, extracted from normal mode splittings vs.\ pump strength, $\xi$, and external flux.
	(f) Observed SNAIL-spin coupling, $\tilde{g}_{ms}^*$, extracted from normal mode splittings, vs.\ pump strength, $\xi$, and external flux.
    }
    \label{fig:parametric_coupling}
\end{figure}

To investigate pump-activated interactions, we next applied a pump tone and probed the cavity response as a function of pump frequency and power [\cref{fig:parametric_coupling}(a)].
In the range where the pump frequency, $f_\text{pump}$, is close to the difference between SNAIL and cavity/spin frequency, two avoided crossings appear~[\cref{fig:parametric_coupling}(b)].
The larger one corresponds to dynamic hybridization between the SNAIL and the cavity, induced when the pump frequency matches their difference.
Setting $\hbar=1$, this interaction can be written as ${\mathcal{H}}_{mc} = \tilde g_{mc} (\hat m^\dagger \hat c + \hat m \hat c^\dagger)$, where the induced coupling is predicted to be $\tilde g_{mc} \approx 6 (g_{mc}/\Delta_{mc})\xi g_3$.
Here $\xi$ is the pump amplitude expressed as a displacement of the coupler mode, $\propto \sqrt{n_p}$, where $n_p$ is the number of steady-state pump photons in the SNAIL mode;
$g_3$ is the third-order nonlinearity coefficient of the SNAIL~\cite{zhou_realizing_2023,chapman_highonoffratio_2023}.
The smaller avoided crossing, offset in frequency, corresponds to dynamic hybridization with the spin ensemble, with ${\mathcal{H}}_{ms} = \tilde g_{ms} (\hat m^\dagger \hat s + \hat m \hat s^\dagger)$.
The cavity coupling is larger because the SNAIL couples directly to the cavity, whereas the spins couple to the SNAIL only through the cavity. 
As a result, the spin-SNAIL coupling is reduced by approximately the static cavity-spin participation, $\tilde g_{ms} \approx \tilde g_{mc} \times (g_{cs}/\Delta_{cs})$~\cite{zhou_realizing_2023}.
This picture quantitatively explains the spectra observed while the pump is applied, as shown by the overlaid simulation in \cref{fig:parametric_coupling}(b).
See \cref{section:Theory} for details.

Next, we investigated the maximally attainable parametric coupling.
We first measured the magnitude of the induced normal-mode splitting between cavity and SNAIL, $2\tilde{g}_{mc}$, as a function of pump amplitude, $\xi$, at several flux values.
At low pump strengths we recover the perturbative prediction of a linear dependence of $\tilde{g}_{mc}$ on $\xi$.
At higher powers, the coupling deviates from this scaling [\cref{fig:parametric_coupling}(c,e)], consistent with previous observations of strongly pumped Josephson couplers \cite{chapman_highonoffratio_2023}.
The pump-power calibration and the coupling fits are described in detail in~\cref{section:Meas}.

Performing the same measurement at the SNAIL-spin difference frequency realizes strong pump-induced coupling to the spins [\cref{fig:parametric_coupling}(d,f)].
The behavior at high pump power, however, is qualitatively different from that of the SNAIL-cavity coupling.
Rather than a gradual deviation from linear scaling, the coupling extracted from the observed normal-mode splitting, $\tilde{g}^*_{ms}$, saturates at a ceiling near $1$~MHz---close in magnitude to the static cavity-spin coupling, $g_{cs}$.
The observed splitting already unambiguously demonstrates strong coupling between SNAIL and spins; below, we further show that it does not reflect the true parametric coupling rate, $\tilde{g}_{ms}$.

\begin{figure}
    \includegraphics[]{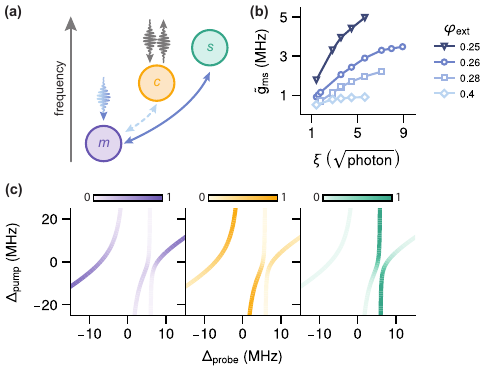}
    \caption{
    \textbf{Suppression of observed parametric spin coupling.}
    (a) Strong pump-induced coupling between SNAIL and spins (solid blue) also induces a detuned coupling to the nearby cavity (dashed light blue).
    (b) True parametric coupling, $\tilde g_{ms}$, inferred from the observed coupling, $\tilde g^*_{ms}$.
    (c) Overlap of each driven eigenmode of the system (as shown spectroscopically in \cref{fig:parametric_coupling}(b)) with the bare SNAIL, cavity, and spin modes, from left to right.
    }
    \label{fig:modeling}
\end{figure}

We find that the saturation of the observed coupling at $g_{cs}$ is a direct consequence of the interactions between all three modes, and of how it is measured in continuous-wave spectroscopy.
Two separate effects are at play.
First, the pump activates not only the targeted SNAIL-spin coupling but also off-resonant SNAIL-cavity couplings [\cref{fig:modeling}(a)].
The total interaction Hamiltonian in this situation is $\tilde{\mathcal{H}}_{mcs} = \tilde g_{mc} (\hat m^\dagger \hat c\, \mathrm e^{-i\Delta_{cs}t} + \mathrm{h.c.}) + \tilde g_{ms} (\hat m^\dagger \hat s + \mathrm{h.c.})$.
Second, the prefactor of the SNAIL-cavity coupling is enhanced over that of the SNAIL-spin coupling by the ratio $\Delta_{cs}/g_{cs}$; as described previously, this is a necessary consequence of the fact that the cavity acts as a bus connecting SNAIL and spin modes.
Combining these effects, the observed coupling can be expressed as $\tilde{g}^*_{ms} = \tilde g_{ms}\cos{\Lambda} \equiv \tilde g_{ms}/(\sqrt{1 + (\tilde{g}_{mc}/\Delta_{cs})^2})$, where the dressing factor $\cos{\Lambda}$ describes the SNAIL-cavity hybridization induced by the off-resonant SNAIL-cavity drive (see \cref{section:Theory}).
At small pump amplitudes, $\cos{\Lambda} \approx 1$ and the splitting grows linearly with $\xi$.
As $\xi$ increases, however, $\tilde g_{ms}$ grows linearly while $\cos{\Lambda}$ scales inversely.
The resulting product asymptotes to the static coupling $g_{cs}$, independent of the pump power.
Together, these effects mean that under a strong pump the spin degree of freedom is not only hybridized with the SNAIL but also leaks into the cavity.
A decomposition of the simulated eigenmodes into the bare modes confirms this picture [\cref{fig:modeling}(c)]: 
at the operating point all three modes contribute appreciably to each polariton.
To recover the parametric coupling strength --- the magnitude $\tilde g_{ms}$ entering the interaction Hamiltonian --- we divide out the apparent suppression from the cavity dressing [\cref{fig:modeling}(b)].
This analysis restores a behavior reminiscent of the SNAIL-cavity coupling: a steady deviation from linear scaling at high pump power, but no hard ceiling.
Noteably, we achieve a parametric coupling strength that exceeds the static coupling of the spins to rest of the system.

The distinction between $\tilde g_{ms}$ and $\tilde g^*_{ms}$ has direct consequences for state transfer.
At first glance, the cavity dressing appears to limit efficient swap rates to $\ll$~1~MHz because of the bus mode's strong participation at large $\xi$.
However, it is well-known that leakage into close-by modes can be avoided using time-varying, rather than continuous-wave drives~\cite{motzoi_simple_2009}.
Indeed, numerical simulations using the experimentally extracted parameters together with only a simple pulse-shaping protocol predict that a single excitation can be transferred between SNAIL and spins with $\gtrsim 98\%$ efficiency in $\sim 200$~ns.
This suggests that the speed limit for swapping excitations between circuit and spins will be set by $\tilde g_{ms}$ and $\Delta_{cs}$, rather than by $\tilde g^*_{ms}$, in agreement with existing models for limits on state leakage~\cite{zhu_quantum_2021}.
Importantly, the amplitude and phase modulation required for pulse shaping are easily implemented with SNAIL couplers that use parametric microwave drives~\cite{zhou_realizing_2023}.
The saturation observed in spectroscopy therefore does not preclude fast, faithful state transfer in our architecture.
See \Cref{section:Theory} for details on analysis and modeling.

In summary, we have demonstrated a dynamically controlled beamsplitter interaction between a superconducting quantum circuit and a solid-state spin ensemble.
We have realized an architecture that integrates a SNAIL three-wave mixer with a cavity-based hybrid device. 
We then used parametric pumping to achieve on-demand strong coupling between the circuit and a collective $^{171}\mathrm{Yb}^{3+}$ spin excitation, evidenced by a resolved normal-mode splitting.
Despite the potential for information leakage into the bus mode, conventional pulse-shaping techniques will enable quantum state transfer between circuit and spins on timescales well below 1~\textmu s.

A natural next step is the integration of a qubit into this architecture~\cite{kubo_hybrid_2011,zhu_coherent_2011};
a transmon qubit, for example, could easily be accommodated on the same chip as the SNAIL coupler.
The principal remaining challenges for high-fidelity quantum state transfer are then inhomogeneous broadening and spin reset.
Multiple established and compatible avenues to tackle them exist.
Inhomogeneity can be overcome with adiabatic pulse shaping to render state transfer insensitive to line broadening~\cite{osullivan_randomaccess_2022}; 
efficiency can further be increased using ensembles with substantially narrower inhomogeneous linewidths, such as $^{171}\mathrm{Yb}^{3+}$:CaWO$_4$~\cite{tiranov_subsecond_2025} or using optimized coupler designs that may increase dynamic coupling~\cite{maiti_linear_2025b}.
Spin reset is a practical concern because of long spin relaxation times; in the spin system used here, several hours were reported~\cite{chiossi_optical_2025}. 
In the present experiment, no significant spin population was generated, and reset was thus not an important concern.
However, to make our architecture compatible with control protocols that create a large spin population~\cite{julsgaard_quantum_2013}, we can employ resonators that enable fast reset via Purcell-relaxation~\cite{bienfait_controlling_2016}.
Combining these techniques with spin echoes~\cite{ranjan_multimode_2020} or atomic frequency combs~\cite{putz_spectral_2017} will then enable an efficient, multi-mode spin ensemble memory.

Our results thus establish parametric coupling as a promising avenue for high-fidelity quantum interfaces between superconducting circuits and solid-state spin ensembles.
Combined with refocusing protocols, state-of-the-art Josephson couplers, and qubits, our architecture will enable quantum memories that can far exceed those available within superconducting circuits alone.
The ability to tailor the interactions between circuits and spins additionally presents new opportunities for the investigation of many-body phenomena~\cite{lei_manybody_2023}, metrological applications such as collective spin squeezing~\cite{groszkowski_reservoirengineered_2022}, and microwave-to-optical transduction~\cite{xie_scalable_2025}.

\paragraph{Acknowledgements ---} 
This research was carried out in part in the Materials Research Lab Central Facilities and the Holonyak Micro and Nanotechnology Lab at the University of Illinois.
We thank K.~Chow and R.~Gon\c{c}alves for help with SNAIL device fabrication, E.~Goldschmidt for providing the YSO crystal, and A.~Kou, C.~Lang, and P.~Bertet for critical reading of the manuscript draft.
Design, fabrication, experimental data acquisition and analysis was supported by the National Science Foundation under awards no.~2016136 and no.~2137642. 
Theoretical modeling of resonator-spin interactions was additionally supported by the U.S.~Department of Energy under contract no.~DE-SC0022060.
A.B.\ acknowledges support from the Alfred P. Sloan Minority Ph.D.\ Program, from a SURGE fellowship by the University of Illinois, and from a Graduate College Fellowship by the University of Illinois.
S.R.\ acknowledges support from a Charles P.~Slichter Fellowship by the University of Illinois. 

\appendix

\section{Experimental Setup} \label{section:exp_set}

\subsection{Device Design and Manufacturing} \label{ssec:dm}

The SNAIL was fabricated as follows.
A 200~nm NbTiN film was sputtered onto a 2-inch, 430~$\mu$m thick sapphire wafer (University Wafers) in a Plassys MEB550S4II UHV system.
SNAIL capacitor pads were defined by photolithography with a Heidelberg MLA150, descummed in O$_2$ plasma (2~min, 300~mTorr, 18~W), and etched with O$_2$/SF$_6$ in an Oxford ICP-RIE.
Josephson junctions were patterned by e-beam lithography on an Elionix 150~keV system and formed via the bridge-free technique using double-angle aluminum evaporation at $\pm 45^{\circ}$ and 0.3~nm/s, with an argon mill prior to the first deposition, an 80-min, 60-Torr oxidation between depositions, and a 5-min, 10-Torr capping oxidation.
Excess aluminum was removed via overnight liftoff in NMP at 80$^{\circ}$C.
Chips were then coated with photoresist for protection, diced on an ADT 7122 wafer dicer, cleaned, probed at room temperature to verify junction parameters, and clamped into the cavity for cooldown.

The spin ensemble crystal is a $3 \times 4 \times 5$~mm, 10~ppm isotopically purified $^{171}\mathrm{Yb}^{3+}$:$\mathrm{Y}_2\mathrm{SiO}_5$ crystal grown by the Czochralski method (Teledyne FLIR), cut along the $\mathbf{D_1}$, $\mathbf{D_2}$, and $\mathbf{b}$ polarization-extinction axes.
The faces perpendicular to $\mathbf{b}$ were polished.

The cavity is a 3-loop, 2-gap design previously used for coupling to rare-earth spin ensembles \cite{chen_coupling_2016}; its differential mode produces a uniform, strong magnetic field over the central-loop volume \cite{choi_ultrastrong_2023, goryachev_highcooperativity_2014a} (\cref{fig:supp_cavity}).
The cavity frequency is tuned by inserting a 430~$\mu$m sapphire chip into the 450~$\mu$m gap; we target a room-temperature frequency $\sim 10$~MHz below the desired base-temperature value to compensate for the shift caused by thermal contraction.
We dub the full device the \underline{Dis}persively \underline{Co}upled \underline{Al}uminum \underline{Pa}rametric \underline{Ca}vity (DisCoAlPaCa).

\subsection{Spin Transition Targeting} \label{ssec:spins}

Two properties of $^{171}\mathrm{Yb}^{3+}$:$\mathrm{Y}_2\mathrm{SiO}_5$ are relevant for our experiment: the ground-state transition energies and their coupling to an applied AC magnetic field.
At our base temperature, the ensemble is predominantly in the electronic ground state.
Transition frequencies have been reported elsewhere~\cite{tiranov_spectroscopic_2018}; we summarize the calculation below.

The electronic spin $\mathbf{S}$ ($S=1/2$) and nuclear spin $\mathbf{I}$ ($I=1/2$) are coupled through the hyperfine tensor $\mathbf{A}$, giving
\begin{equation}
    \mathcal{H} = \mathbf{I} \cdot \mathbf{A} \cdot \mathbf{S} + \mu_B \mathbf{B} \cdot \mathbf{g} \cdot \mathbf{S} - \mu_n \mathbf{B} \cdot \mathbf{g}_n \cdot \mathbf{I},
\end{equation}
where $\mathbf{g}$ and $\mathbf{g}_n$ are the electronic and nuclear $g$-tensors, and $\mu_B$ and $\mu_n$ are the Bohr and nuclear magnetons.
The quadrupolar term vanishes for $I=1/2$.
The nuclear interaction is taken isotropic with $g_n = 0.987$, at least an order of magnitude smaller than the effective electronic $g$-factors and typically not resolved.

In our experiment $\mathbf{B} = 0$, so both Zeeman terms vanish.
The remaining Hamiltonian diagonalizes analytically into four levels with energies
\begin{equation}
    \tfrac{1}{4}\left[ -A_3 \pm (A_1 + A_2) \right], \quad
    \tfrac{1}{4}\left[ A_3 \pm (A_1 - A_2) \right].
    \label{eqn:hyperfine_diag}
\end{equation}

For each set of ground-state hyperfine eigenvalues $(A_1^{(g)}, A_2^{(g)}, A_3^{(g)})$ this yields four levels and six transitions; the three independent eigenvalues of $^{171}\mathrm{Yb}^{3+}$:$\mathrm{Y}_2\mathrm{SiO}_5$ make all transitions nondegenerate.
$^{171}\mathrm{Yb}^{3+}$ substitutes for $\mathrm{Y}^{3+}$ at two magnetically inequivalent sites, doubling the set to twelve transitions.
We select one with sufficient energy and electronic $g$-factor to couple to a 3D microwave cavity.

The coupling to an external AC field at a given orientation is set by $\mathbf{g}$.
Computing $\mathbf{g}$ along each polarization-extinction axis gives the magnetic-field sensitivity of every transition for arbitrary applied fields.
We convert this into expected cavity coupling using finite-element simulations of the cavity in Ansys HFSS: integrating the mode magnetic field over the crystal volume and combining with

\begin{equation}
    \bar g = \mu_B g |\vec{B}_{\mathrm{ZPF}}|,
\end{equation}
\begin{equation}
    |\vec{B}_{\mathrm{ZPF}}| = \frac{\mu_0 \int H \, dV}{V} \sqrt{\frac{h \nu}{2 \,(1~\mathrm{J})}},
\end{equation}
\noindent yields the collective coupling $\sqrt{N}\,\bar g$, with $\bar g$ the single-spin coupling (cf.\ main text), $N$ the number of spins available for collective coupling, and $\nu$ the cavity-mode frequency~\cite{angerer_collective_2016}.
We chose the site I transition between $\ket{1}$ and $\ket{3}$ at 2.87~GHz, with gyromagnetic ratio $\gamma_e = \mu_B g/h = 9.7$~GHz/T along $\mathbf{D_2}$ and a predicted coupling of 1.4~MHz; the measured static coupling is 1.1~MHz, with the difference explainable by incomplete ground-state polarization.

\subsection{Device Characterization} \label{ssec:dc}

Device parameters extracted from spectroscopy fits are summarized in \cref{tab:supp_system_parameters}.

\begin{figure}[tb]
    \includegraphics{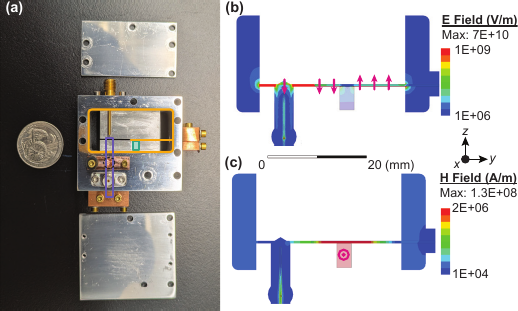}
    \caption{
    \textbf{DisCoAlPaCa Design and Simulation.}
    (a) DisCoAlPaCa assembly with caps and an American quarter for scale (24.257~mm diameter); colored outlines indicate mode elements following the paper's color scheme.
    The tuning dielectric is held in the right cavity gap by a clamp.
    The flux coil is held in a copper cap over the SNAIL chip. Its position is indicated with a black dashed line.
    (b) Simulated electric and magnetic fields of the differential mode; magenta arrows indicate field direction.
    }
    \label{fig:supp_cavity}
\end{figure}

\begin{table}[htb]
    \centering
    \begin{tabular}{|l|c|c|}
    \hline
    \textbf{Bare Mode} & \textbf{Frequency (GHz)} & \textbf{Linewidth (kHz)} \\
    \hline
    Spin & 2.8703 & 460 [$\gamma$] \\
    Cavity & 2.858 & 200 [$\kappa$] \\
    SNAIL & 2.45 -- 2.95 & 800 -- 1400 [$\kappa_m$] \\
    \hline
    \textbf{Coupling} & \textbf{Value (MHz)} & -- \\
    \hline
    $g_{cs}$ & 1.1 & --  \\
    $g_{mc}$ & 27.4 & --  \\
    \hline
    \end{tabular}
    \caption{
    \textbf{System parameters.}
    Mode parameters extracted from microwave-spectroscopy fits at the appropriate flux points.
    }
    \label{tab:supp_system_parameters}
\end{table}

\begin{figure}
    \includegraphics[]{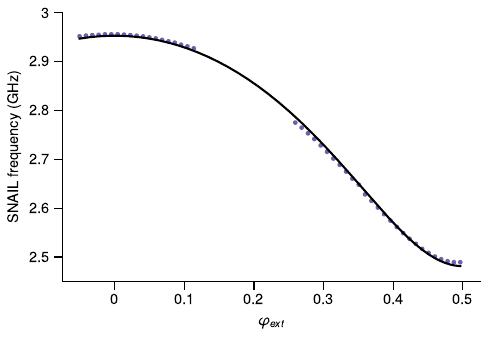}
    \caption{
    \textbf{SNAIL fitting.}
    SNAIL frequency vs.\ external flux, fit by varying the SNAIL inductance.
    Fit parameters are reported in \cref{tab:supp_snail_params}.
    The flux quantum was calibrated by sweeping the coil current over one full period of the SNAIL frequency.
    }
    \label{fig:supp_SNAIL}
\end{figure}

SNAIL parameters are extracted by fitting the flux dependence of the SNAIL frequency (\cref{fig:supp_SNAIL}) to
\begin{align}
    f_{\mathrm{SNAIL}}(\varphi_{\mathrm{ext}})
    &= \frac{1}{2\pi \sqrt{L_{m,\mathrm{tot}}\, C_m}} \nonumber \\
    &= \frac{1}{2\pi \sqrt{[L_{\mathrm{SNAIL}}(\varphi_{\mathrm{ext}}) + L_{\mathrm{lin}}]\, C_m}},
\end{align}
with
\begin{equation}
    L_{\mathrm{SNAIL}}(\varphi_{\mathrm{ext}}) = \frac{L_0}{2 c_2(\varphi_{\mathrm{ext}})}.
\end{equation}
The linear inductance $L_{\mathrm{lin}}$ and capacitance $C_m$ are simulated using Ansys HFSS and pyEPR \cite{minev_energyparticipation_2021}; see \cref{ssec:SNAIL} for the calculation of $c_2$.
Extracted SNAIL parameters are listed in \cref{tab:supp_snail_params}.

\begin{table}[htbp]
    \centering
    \begin{tabular}{|l|l|l|}
        \hline
        \textbf{Parameter} & \textbf{Value} & \textbf{Energy} \\
        \hline
        $C_m$ & 280 fF & 69.2 MHz \\
        $L_{\mathrm{lin}}$ & 4.00 nH & 40.9 GHz \\
        $L_J$ & 2.65 nH & 61.6 GHz \\
        $\alpha$ & 0.084 & -- \\
        \hline
    \end{tabular}
    \caption{
    \textbf{SNAIL parameters.}
    Parameters extracted from the fit in \cref{fig:supp_SNAIL}.
    }
    \label{tab:supp_snail_params}
\end{table}

\subsection{Cryogenic Measurement Setup} \label{ssec:cryo}

\begin{figure}[]
    \includegraphics[]{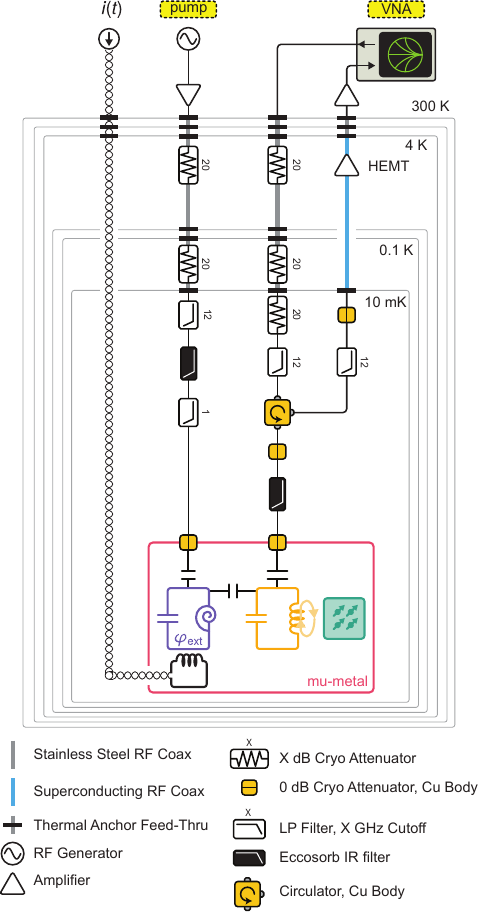}
    \caption{
    \textbf{Experimental setup and wiring diagram.}
    Components: low-frequency circulator (Low Noise Factory LNF-CIC2\_2.8A), low-pass filters (K\&L 6L250-12000/T26000), additional low-pass filter on the SNAIL pump line (Mini-Circuits VLF-800+), HEMT amplifier (LNF-LNC0.3\_14B), room-temperature amplifier (LNF-LNR1\_15B\_5V).
    The SNAIL pump line is attenuated by $-40$~dB and the cavity probe line by $-60$~dB.
    Output filtering is minimal because of the limited availability of isolators in the spin frequency band.
    The flux coil is thermalized at the 4~K plate and driven directly by a Yokogawa GS210 current supply.
    Pumps are gated using a Rohde \& Schwarz SGS100A.
    Spectroscopy is performed with a Keysight P9374A VNA; time-domain measurements and some basic spectroscopy use a Xilinx FPGA board (ZCU216) with QICK firmware~\cite{stefanazzi_qick_2022a} in place of the VNA.
    Transmission ($S_{21}$) data were acquired through the SNAIL port during a cooldown without the 1~GHz low-pass filter on the output.
    }
    \label{fig:supp_wiring}
\end{figure}

The device was cooled to 10~mK in an Oxford Instruments Triton 500 dilution refrigerator.
The wiring is shown in \cref{fig:supp_wiring}.

\section{Theoretical Modeling} \label{section:Theory}

\subsection{The SNAIL Hamiltonian} \label{ssec:SNAIL}

We outline the SNAIL Hamiltonian to define the nonlinearities used in the parametric coupling, following \cite{frattini_3wave_2017, frattini_threewave_2021, frattini_optimizing_2018, chapman_highonoffratio_2023, zhou_realizing_2023}.

The SNAIL is an asymmetric loop containing three large Josephson junctions on one arm and one smaller junction on the other; an external flux tunes its inductance and nonlinearities.
Expanded around an energy-minimum flux point, the inductive potential supports three-wave mixing, which mediates parametrically tunable coupling between two modes.

The SNAIL potential at fixed external flux is
\begin{equation}
    \frac{U_s(\hat{\varphi}_s)}{E_J}
    = -\alpha \cos \hat{\varphi}_s
    - 3 \cos \frac{\hat{\varphi}_s - \varphi_\mathrm{ext}}{3},
\end{equation}
where $\varphi_\mathrm{ext} \equiv \Phi_\mathrm{ext}/\Phi_0$ is the reduced external flux, $\hat{\varphi}_s$ is the phase across the SNAIL, $\alpha = E_{J,\mathrm{small}}/E_{J,\mathrm{large}} < 1$, and $E_J$ denotes the inductive energy of a single large junction.

The minimum $\varphi_{\mathrm{min}}$ satisfies
\begin{equation}
    \alpha \sin \hat{\varphi}_\mathrm{min} - \sin \frac{\hat{\varphi}_\mathrm{min} - \varphi_\mathrm{ext}}{3} = 0,
\end{equation}
which has a single solution for $\alpha < 1/3$.
We target $\alpha \approx 0.1$ to suppress flux noise sensitivity.
Near $\varphi_{\mathrm{min}}$ the potential expands as
\begin{equation}
    U_s \approx \sum_{n=2}^{6} \frac{c_n}{n!}(\hat{\varphi}_s - \varphi_{\mathrm{min}})^n + \mathcal{O}(\hat{\varphi}_s^7),
\end{equation}
with coefficients
\begin{align}
    c_2 &= \alpha \cos \hat{\varphi}_s + \tfrac{1}{3} \cos \tfrac{\hat{\varphi}_s - \varphi_\mathrm{ext}}{3}, \\
    c_3 &= -\alpha \sin \hat{\varphi}_s - \tfrac{1}{9} \sin \tfrac{\hat{\varphi}_s - \varphi_\mathrm{ext}}{3}, \\
    c_4 &= -\alpha \cos \hat{\varphi}_s - \tfrac{1}{27} \cos \tfrac{\hat{\varphi}_s - \varphi_\mathrm{ext}}{3}, \\
    c_5 &= \alpha \sin \hat{\varphi}_s + \tfrac{1}{81} \sin \tfrac{\hat{\varphi}_s - \varphi_\mathrm{ext}}{3}, \\
    c_6 &= \alpha \cos \hat{\varphi}_s + \tfrac{1}{243} \cos \tfrac{\hat{\varphi}_s - \varphi_\mathrm{ext}}{3}.
\end{align}

Including the linear inductance and shunt capacitance of the full circuit rescales the coefficients $c_j \to \tilde{c}_j$ without changing the location of the minima \cite{frattini_threewave_2021}:
\begin{align}
    \tilde{c}_2 &= p\, c_2, \\
    \tilde{c}_3 &= p^3 c_3, \\
    \tilde{c}_4 &= p^4 \!\left[ c_4 - \tfrac{3 c_3^2}{c_2} (1-p) \right], \\
    \tilde{c}_5 &= p^5 \!\left[ c_5 - \tfrac{10 c_4 c_3}{c_2} (1-p) + \tfrac{15 c_3^3}{c_2^2} (1-p)^2 \right], \\
    \tilde{c}_6 &= p^6 \Big[ c_6 - \tfrac{10 c_4^2 + 15 c_5 c_3}{c_2}(1-p) \nonumber \\
        &\qquad\quad + \tfrac{105 c_4 c_3^2}{c_2^2}(1-p)^2 - \tfrac{105 c_3^4}{c_2^3}(1-p)^3 \Big],
\end{align}
where $p = L_s/(L + L_s)$ is the inductive participation ratio, $L_s = L_J/c_2$, and $L$ is the linear inductance.

After second quantization with $\hat{\varphi} - \varphi_{\mathrm{min}} = \phi_m(\hat{m} + \hat{m}^\dagger)$, $\hat{N} = -i \phi_m^{-1}(\hat{m} - \hat{m}^\dagger)/2$, and zero-point fluctuations $\phi_m = (2 E_C / (\tilde{c}_2 E_J))^{1/4}$, the SNAIL Hamiltonian splits into a linear and a nonlinear part:
\begin{equation}
    \hat{\mathcal{H}}_m = \omega_m \hat{m}^\dagger \hat{m} + \hat{\mathcal{H}}_{mnl},
\end{equation}
\begin{align}
    \hat{\mathcal{H}}_{mnl}
    = \sum_{n=3}^{6} g_n (\hat{m} + \hat{m}^\dagger)^n + \mathcal{O}(\hat{m}^7),
\end{align}
with
\begin{align}
    \omega_m &= \sqrt{8 \tilde{c}_2 E_C E_J}, \\
    g_n &= E_J\, \phi_m^n\, \frac{\tilde{c}_n}{n!} \quad (n = 3,4,5,6).
    \label{supp_eqn_g_3}
\end{align}

\subsection{The Coupler Static and Driven Hamiltonian} \label{ssec:coupler_stat_driv}

The cavity $\hat{c}$ and spin ensemble $\hat{s}$ are dispersively coupled with strength $g_{cs} = \sqrt{N}\,\bar g$ and detuning $\Delta_{cs}$.
The spin ensemble bright mode $\hat{s} = (\sum_j g_j \sigma_-^{(j)})/\bar g\sqrt{N}$ defined in the main text is treated as a harmonic oscillator via the Holstein-Primakoff approximation \cite{diniz_strongly_2011}, valid at low excitations.
The cavity is additionally coupled to the SNAIL $\hat{m}$ with strength $g_{mc}$ and detuning $\Delta_{mc}$.
The bare Hamiltonian is
\begin{align}
    \hat{\mathcal{H}}
    &= \omega_c \hat{c}^\dagger \hat{c} + \omega_s \hat{s}^\dagger \hat{s} + \omega_m \hat{m}^\dagger \hat{m} + \hat{\mathcal{H}}_{mnl} \nonumber \\
    &\quad - g_{mc}(\hat{m} - \hat{m}^\dagger)(\hat{c} - \hat{c}^\dagger)
    - g_{cs}(\hat{c} - \hat{c}^\dagger)(\hat{s} - \hat{s}^\dagger).
\end{align}

In the rotating-wave approximation,
\begin{align}
    \label{eqn:supp_coupled_RWA}
    \hat{\mathcal{H}}
    &= \omega_c \hat{c}^\dagger \hat{c} + \omega_s \hat{s}^\dagger \hat{s} + \omega_m \hat{m}^\dagger \hat{m} + \hat{\mathcal{H}}_{mnl} \nonumber \\
    &\quad + g_{mc}(\hat{m} \hat{c}^\dagger + \hat{m}^\dagger \hat{c}) + g_{cs}(\hat{c} \hat{s}^\dagger + \hat{c}^\dagger \hat{s}).
\end{align}

For the driven, renormalized system,
\begin{equation}
    \hat{\mathcal{H}}
    = \omega_c \hat{c}^\dagger \hat{c} + \omega_s \hat{s}^\dagger \hat{s} + \omega_m \hat{m}^\dagger \hat{m}
    + \hat{\mathcal{H}}_d + \hat{\mathcal{H}}_{mnl},
\end{equation}
where $\hat{\mathcal{H}}_d$ is the drive.
Because all three modes are hybridized, the SNAIL flux operator carries contributions from each:
$\hat{\varphi}_m - \varphi_{\mathrm{min}} = \phi_m(\hat{m} + \hat{m}^\dagger) + \phi_c(\hat{c} + \hat{c}^\dagger) + \phi_s(\hat{s} + \hat{s}^\dagger)$, with
$\phi_c \approx \phi_m\, g_{mc}/\Delta_{mc}$ and
$\phi_s \approx \phi_m\, g_{mc} g_{cs}/(\Delta_{mc} \Delta_{cs})$ in the dispersive regime \cite{blais_circuit_2021}.
The extra factor in $\phi_s$ reflects the concatenated coupling: the spins inherit mixer participation only through the cavity.

The drive at frequency $\omega_p$ and amplitude $\epsilon$ is
\begin{equation}
    \hat{\mathcal{H}}_d = -(\epsilon e^{-i\omega_p t} - \epsilon^* e^{i\omega_p t})(\hat{m} - \hat{m}^\dagger).
\end{equation}
Moving to the displaced frame~\cite{chapman_highonoffratio_2023} with $\hat{U}_d = \exp[\tilde{\xi}(t) \hat{m}^\dagger - \tilde{\xi}^*(t) \hat{m}]$ and $\tilde{\xi} \equiv \xi e^{-i\omega_p t} \equiv [\epsilon/(\omega_p - \omega_m)] e^{-i\omega_p t}$, then into the co-rotating frame and dropping counter-rotating terms and coupler dissipation, the flux operator becomes
\begin{align}
    \hat{\tilde{\varphi}}
    &= \phi_c(\hat{c} e^{-i\omega_c t} + \hat{c}^\dagger e^{i\omega_c t})
    + \phi_s(\hat{s} e^{-i\omega_s t} + \hat{s}^\dagger e^{i\omega_s t}) \nonumber \\
    &\quad + \phi_m(\hat{m} e^{-i\omega_m t} + \hat{m}^\dagger e^{i\omega_m t} + \xi e^{-i\omega_p t} + \xi^* e^{i\omega_p t}).
\end{align}

When $\omega_p$ is detuned by $\Delta_\mathrm{pump} = \omega_p - \Delta_{mc}$ from the mixer-cavity difference frequency, a beamsplitter interaction
\begin{equation}
    \label{eqn:supp_bs}
    g_{\mathrm{bs}}\, e^{i\theta} \hat{m}^\dagger \hat{c}\, e^{i t \Delta_\mathrm{pump} } + \mathrm{h.c.}
\end{equation}
emerges, with leading-order rate
\begin{align}
    \tilde{g}_{mc} \equiv g_{\mathrm{bs}} &\approx E_J\, \phi_c \phi_m^2\, \xi\, \tilde{c}_3
    \approx 6 \,\frac{g_{mc}}{\Delta_{mc}}\, \xi\, g_3.
\end{align}

The same form holds for the mixer-spin interaction with an additional factor $g_{cs}/\Delta_{cs}$.
This expression is leading-order in pump amplitude; higher-order corrections become important at strong drives, producing nonlinear deviations from the linear $\xi$ scaling \cite{chapman_highonoffratio_2023, baskov_exact_2025b}.
A complete description of strongly pumped Josephson couplers remains an active research direction, with Floquet methods \cite{xia_exceeding_2025} required to capture transitions in strongly driven nonlinear circuits.

\subsection{Cavity-Mediated Renormalization of the Pumped Coupling} \label{ssec:parasitic}

We consider the driven mixer simultaneously parametrically coupled to the cavity and to the spins.
To explain our spectroscopy data, we transform the frame of the SNAIL to rotate at $-\omega_p$.
After the RWA,
\begin{align}
    \hat{\mathcal{H}}
    &= \omega_s \hat{s}^\dagger \hat{s} + \omega_c \hat{c}^\dagger \hat{c} + (\omega_m + \omega_p) \hat{m}^\dagger \hat{m} \nonumber \\
    &\quad + \tilde{g}_{ms}(\hat{m} \hat{s}^\dagger + \hat{m}^\dagger \hat{s}) + \tilde{g}_{mc}(\hat{m} \hat{c}^\dagger + \hat{m}^\dagger \hat{c}).
\end{align}
It is intuitive that this is the correct frame to describe our spectroscopy data: for $\omega_p = \omega_{c(s)} - \omega_m$, the SNAIL is parametrically brought on resonance with the cavity (spins).
This is the Hamiltonian diagonalized to produce the three-mode simulation in \cref{fig:parametric_coupling}(b).

The observed coupling in \cref{fig:parametric_coupling}(f) is between a hybridized mixer-cavity mode and the spins. 
We enter a frame that considers the hybridization between the mixer and the cavity to get a better understanding of the observed spin coupling. 
We rediagonalize the system in the basis of the off-resonantly coupled mixer-cavity subsystem with a Bogoliubov transformation.
With $\Delta_{cs} = \omega_s - \omega_c$ and $\Delta = \omega_m + \omega_p - \omega_c$ and applying a Bogoliubov transformation with $\Lambda = \tfrac{1}{2}\arctan(2 \tilde{g}_{mc}/\Delta)$ produces dressed mixer/cavity frequencies
\begin{align}
    \tilde{\omega}_m &= \tfrac{1}{2}\!\left( \omega_m + \omega_p + \omega_c + \sqrt{\Delta^2 + 4 \tilde{g}_{mc}^2} \right), \\
    \tilde{\omega}_c &= \tfrac{1}{2}\!\left( \omega_m + \omega_p + \omega_c - \sqrt{\Delta^2 + 4 \tilde{g}_{mc}^2} \right),
\end{align}
and the renormalized Hamiltonian
\begin{align}
    \hat{\mathcal{H}}
    &= \omega_s \hat{s}^\dagger \hat{s} + \tilde{\omega}_c \hat{c}^\dagger \hat{c} + \tilde{\omega}_m \hat{m}^\dagger \hat{m} \nonumber \\
    &\quad + \tilde{g}_{ms} \cos\Lambda\, (\hat{m} \hat{s}^\dagger + \hat{m}^\dagger \hat{s})
    + \tilde{g}_{ms} \sin\Lambda\, (\hat{c} \hat{s}^\dagger + \hat{c}^\dagger \hat{s}).
\end{align}

Setting $\tilde{\omega}_m = \omega_s$ (i.e.\ $\Delta = \Delta_{cs} - g_{mc}^2/\Delta_{cs}$) gives the observed mixer-spin coupling
\begin{align}
    \label{eqn:supp_suppression}
    \tilde{g}_{ms}^* = \tilde{g}_{ms} \cos\Lambda
    = \frac{\tilde{g}_{ms}}{\sqrt{1 + (\tilde{g}_{mc}/\Delta_{cs})^2}}.
\end{align}
Dividing the data in \cref{fig:modeling}(b) by $1/\sqrt{1 + (\tilde{g}_{mc}/\Delta_{cs})^2}$ recovers the bare pumped coupling.

\subsection{Simulating Swaps to the Spin Ensemble} \label{ssec:spin_swaps}

\begin{figure}i 
    \includegraphics[]{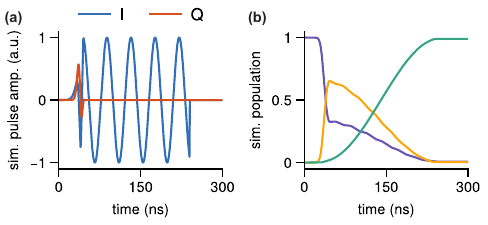}
    \caption{
    \textbf{Simulating swaps to the spins.}
    (a) DRAG pulse used for the simulated swap.
    (b) Time-domain populations of mixer, cavity, and spins during a single-excitation transfer using the pulse in (a).
    The simulation includes coupling and loss rates extracted from the experiment; the final spin population is 98.9\% of the initial mixer population.
    }
    \label{fig:supp_swap_sims}
\end{figure}

To verify that leakage to the cavity mode can be overcome by pulse shaping, we perform QuTiP time-domain simulations of a single-photon population transfer between the mixer and the spin ensemble (\cref{fig:supp_swap_sims}), using couplings and lifetimes extracted from the experiment.
With a simple and not fully-optimized pulse, a single-excitation swap is achieved in 200~ns with 98.9\% final population.

\section{Additional Measurements and Calibrations} \label{section:Meas}

\subsection{Spin Transition Confirmation by repeated cooldowns} \label{ssec:caps}

Without direct spin-ensemble control, we verified the spin mode at 2.87~GHz by repeated cooldowns.
Across five fridge cycles, the transition appeared at $2.87034 \pm 0.00005$~GHz with coupling $1.0 \pm 0.1$~MHz.

\subsection{Spin Transition Confirmation in the ``Halves'' Cavity} \label{ssec:halves}

\begin{figure}
    \includegraphics[]{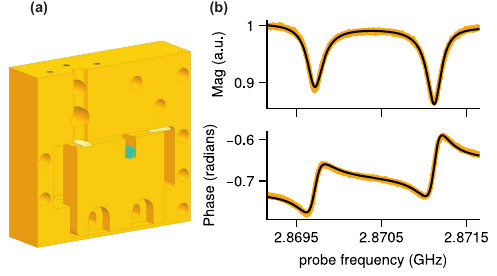}
    \caption{
    \textbf{Halves cavity and spin coupling.}
    (a) Cross section of the alternate cavity, in the same color scheme used elsewhere; no SNAIL is integrated.
    (b) First measurement and fit on this cavity, giving a spin frequency of 2.8704~GHz and a coupling of 710~kHz.
    }
    \label{fig:supp_halves}
\end{figure}

An earlier cavity prototype consisted of two halves and an insert (\cref{fig:supp_halves}), in contrast to the central-body-plus-caps geometry of \cref{fig:supp_cavity}; we refer to them as the ``halves'' and ``caps'' cavities.
A second 10-ppm $^{171}\mathrm{Yb}^{3+}$:$\mathrm{Y}_2\mathrm{SiO}_5$ crystal (cylindrical, 3~mm diameter $\times$ 6~mm height along the crystal $\mathbf{D_2}$ axis) was placed in the halves cavity.

The smaller volume reduces $N$ to about 70\% of the caps-cavity crystal; since the collective coupling $\sqrt{N}\,\bar g \propto \sqrt{N}$, this implies a maximum coupling $\sim 900$~kHz, accounting for incomplete polarization.

The halves cavity was measured without a SNAIL across four cooldowns, yielding an average spin frequency of $2.87031 \pm 0.00005$~GHz and coupling $600 \pm 200$~kHz, with a peak of 746~kHz.
Most measurements clustered between 700 and 750~kHz, with a single 330~kHz outlier likely caused by run-to-run variation in the assembly gap, which sets the field over the crystal volume.

The two crystals give consistent transition frequencies and couplings that scale with crystal size, and the halves measurements were taken without a SNAIL, ruling out a SNAIL-correlated TLS as the origin of the 2.87~GHz feature.

\subsection{Measuring the Pumped Cavity Protection Effect} \label{ssec:CPE}

\begin{figure}
    \includegraphics[]{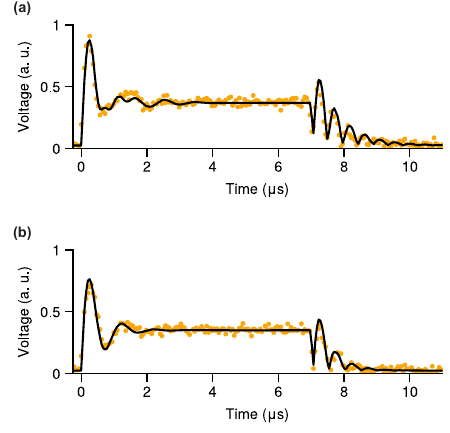}
    \caption{
    \textbf{Static and pumped cavity protection effect.}
    (a) Flux-tuned coupling on resonance with the spin ensemble, reproducing \cref{fig:flux_spec}(e).
    (b) Pump-induced coupling.
    Simulations in both panels use the non-Lorentzian-distribution model of \cite{krimer_nonmarkovian_2014}.
    }
    \label{fig:supp_CPE}
\end{figure}

Following \cite{putz_protecting_2014, krimer_nonmarkovian_2014}, we use the time-domain ring-down of the hybridized system after the drive is switched off as a signature of coupling to a non-Markovian bath: an inhomogeneously broadened ensemble produces a characteristic overshoot and Gaussian decay envelope.
We observe this signature both for static (flux-tuned, \cref{fig:flux_spec}(e), \cref{fig:supp_CPE}(a)) and pump-induced (\cref{fig:supp_CPE}(b)) hybridization.

\subsection{Background Extraction} \label{ssec:BG_removal}

\begin{figure}
    \includegraphics[]{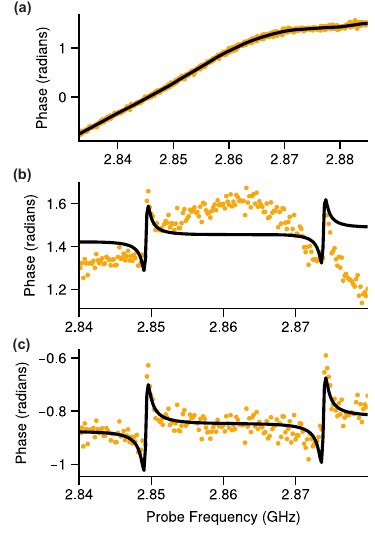}
    \caption{
    \textbf{Background fitting and removal for improved coupling fits.}
    (a) Raw cavity-probe trace from flux spectroscopy of the SNAIL near resonance with the cavity, showing the nonlinear background; black line is the spline fit.
    (b) Phase of a separate mixer-cavity pumped-spectroscopy trace; the uneven background obscures the coupling features.
    (c) Same data as (b) divided by the spline background; the residual phase is approximately linear in probe frequency, simplifying fits and reducing parameter errors.
    }
    \label{fig:supp_BG_removal}
\end{figure}

Pumped-coupling fits are improved by dividing out a nonlinear background extracted from a reference flux-spectroscopy trace.
The reference was taken at 17.5~mA coil current ($\varphi_{\mathrm{ext}} = 0.198$), where the SNAIL is nearly resonant with the cavity.
The pumped-coupling features lie within the avoided crossing of this reference, so we spline-fit the region between the two avoided-crossing peaks and divide the raw data by the result.
This leaves at most a linear phase variation in frequency and improves fit consistency (\cref{fig:supp_BG_removal}).

\subsection{Coupling Fits} \label{ssec:coupling_fit}

We fit pumped and flux-spectroscopy data using two-mode coupled-oscillator forms.
For reflection and transmission respectively
\begin{align}
    S_{11}(\omega) &= A \!\left( 1 + m_t \tfrac{\Delta_c}{\omega_c} \right) e^{-i[\phi_0 + m_\phi(\Delta_c)]} \nonumber \\
    &\quad \times \!\left( 1 + \frac{i \kappa_e (\Delta_s - \frac{i\gamma}{2})}{(\Delta_c - \frac{i(\kappa + \kappa_e)}{2})(\Delta_s - \frac{i\gamma}{2}) - g^2} \right);
\end{align}

\begin{align}
    S_{21}(\omega) &= A \!\left( 1 + m_t \tfrac{\Delta_c}{\omega_c} \right) e^{-i[\phi_0 + m_\phi(\Delta_c)]} \nonumber \\
    &\quad \times \frac{\kappa_e}{i(\Delta_c - \frac{i(\kappa + \kappa_e)}{2})(\Delta_s - \frac{i\gamma}{2}) - g^2}.
\end{align}

Here $\omega_c$ and $\omega_s$ denote the bright and dark mode frequencies, $\Delta_c = \omega - \omega_c$ and $\Delta_s = \omega - \omega_s$ are the detunings between the probe signal and the respective mode frequency, $\kappa$ and $\kappa_e$ the bright-mode internal and external loss rates, $\gamma$ the dark-mode internal loss rate, and $g$ the inter-mode coupling.
The prefactor with $A$, $m_t$, $\phi_0$, $m_\phi$ accounts for amplitude/phase offsets and slopes in the measurement chain.
The remainder is the standard input-output response for two coupled Lorentzian modes \cite{clerk_introduction_2010a, gardiner_input_1985}.
Fits are performed using \texttt{lmfit}.

For SNAIL flux spectroscopy without spin coupling, the bare cavity frequency (extracted from a single-Lorentzian fit at maximum SNAIL detuning) is held fixed and the SNAIL frequency is left free.
At the flux point where the spins are resonant with the hybridized cavity-SNAIL system, the latter plays the role of the bright mode and the spins of the dark mode, with both frequencies free; spin-linewidth initialization uses values from \cite{tiranov_spectroscopic_2018}.
For mixer-cavity pumped fits, the cavity frequency is held fixed.
For mixer-spin pumped fits, both modes are initialized at the spin frequency from flux spectroscopy and left free.

\subsection{Pump-Strength Calibration} \label{ssec:pump_cal}

\begin{figure}
    \includegraphics[]{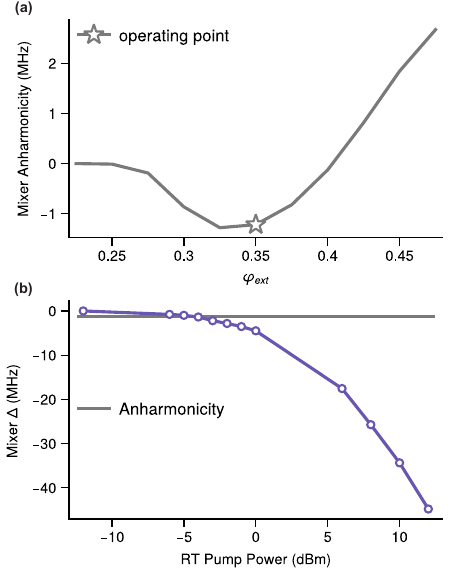}
    \caption{
    \textbf{Pump-strength calibration.}
    (a) Simulated SNAIL anharmonicity (\cref{eq:supp_anharm}) vs.\ $\varphi_{\mathrm{ext}}$; the star marks the operating point used in (b).
    (b) Kerr shift of the SNAIL extracted from pumped cavity-mixer coupling: as the pump strength increases, the SNAIL frequency shifts from its low-power baseline.
    Matching the shift to the simulated anharmonicity sets the single-photon pump power, which we use to convert applied power into pump photons; we then define $\xi$ as the square root of this photon number, the pump-strength coefficient appearing in \cref{eqn:supp_bs}.
    }
    \label{fig:supp_xi_cal}
\end{figure}

We calibrate the pump in mixer-mode photons by matching the measured SNAIL Kerr shift to its simulated anharmonicity, which is given by \cite{frattini_threewave_2021}
\begin{align}
    \label{eq:supp_anharm}
    2 K &= p^3 \!\left[ c_4 - \tfrac{3 c_3^2}{c_2}(1-p) - \tfrac{5}{3}\tfrac{c_3^2}{c_2} p \right]\!\frac{1}{c_2} E_C \nonumber \\
    &\quad + \mathcal{O}\!\big(\omega_a (p \varphi_{\mathrm{zpf}})^4\big).
\end{align}
HFSS eigenmode simulations and pyEPR post-processing \cite{minev_energyparticipation_2021} provide the Hamiltonian parameters needed to evaluate $K$.
\Cref{fig:supp_xi_cal} illustrates the calibration at $\varphi_{\mathrm{ext}} = 0.35$.

\subsection{Extracting $g_3$ from Pumped Coupling} \label{ssec:g3}

\begin{figure}
    \includegraphics[]{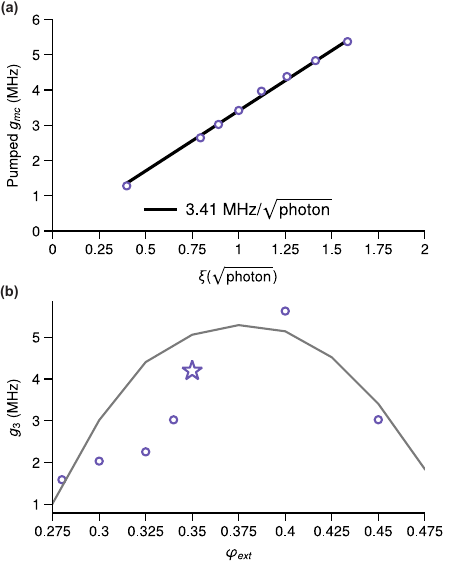}
    \caption{
    \textbf{Mapping $g_3$ vs.\ flux.}
    (a) Low-power data of the pumped mixer-cavity coupling at $\varphi_{\mathrm{ext}} = 0.35$.
    (b) $g_3$ extracted by dividing the measured slope $d g_{\mathrm{bs}}/d\xi$ by the prefactor $6 g_{mc}/\Delta_{mc}$ (\cref{eqn:supp_bs}).
    }
    \label{fig:supp_g3}
\end{figure}

The SNAIL three-wave nonlinearity $g_3$ (\cref{supp_eqn_g_3}) is extracted by fitting the slope of $g_{\mathrm{bs}}$ vs.\ $\xi$ in the linear, low-power regime and dividing by the prefactor $6 g_{mc}/\Delta_{mc}$ from \cref{eqn:supp_bs}.
The result tracks the simulated $g_3$ across flux, providing a qualitative cross-check between measurement and simulation.

% \bibliography{references}

\end{document}